\documentclass[11pt,a4paper]{article}

\usepackage{tikz}
\usetikzlibrary{shapes}
\usepackage{amsmath,amssymb}
\usepackage[T1]{fontenc}
\usepackage{colortbl}
\usepackage{graphicx}
\usepackage[english]{babel}

\definecolor{violet}{rgb}{0.6,0.2,0.6}
\definecolor{brique}{rgb}{.85,.25,0}
\definecolor{dgreen}{rgb}{0,.55,0}
\definecolor{marron}{rgb}{.4,.2,0}
\definecolor{ddcyan}{rgb}{0,.6,.6}
\definecolor{dviolet}{rgb}{.6,0,.6}
 \definecolor{blvert}{rgb}{0,.5,.75}
\definecolor{vertcl}{rgb}{.75,.93,.47}
\definecolor{bleucl}{rgb}{.58,.68,.93}
\definecolor{dcyan}{rgb}{0,.8,.8}
\definecolor{lyellow}{cmyk}{0,0,.2,0}
\definecolor{chair}{rgb}{.9,.7,.6}
\definecolor{rose}{rgb}{.95,.67,.79}
\definecolor{lviolet}{rgb}{.9,0,.6}

\newcommand\rouge[1]{\textcolor{red}{#1}}

\newcommand{\bea}{\begin{eqnarray}}
\newcommand{\eea}{\end{eqnarray}}
\newcommand{\beano}{\begin{eqnarray*}}
\newcommand{\eeano}{\end{eqnarray*}}
\newcommand{\beq}{\begin{equation}}
\newcommand{\eeq}{\end{equation}}

\newcommand{\hs}[1]{\hspace{#1 mm}}

\newcommand{\eps}{\epsilon}

\newcommand{\vph}{\varphi}

    \def\cH{{\cal H}}

\def\fh{{\mathfrak h}}

\def\ft{{\mathfrak t}}

\newcommand{\CC}{{\mathbb C}}

\newcommand{\II}{{\mathbb I}}

\newcommand{\wh}[1]{\widehat{#1}}

\newcommand{\mb}[1]{\hs{4}\mbox{#1}\hs{4}}

\newcommand{\half}{\frac{1}{2}}

\newcommand{\atopn}[2]{\genfrac{}{}{0pt}{}{#1}{#2}}
\newcommand{\up}{\uparrow}
\newcommand{\down}{\downarrow}

\begin{document}

%%%%%%%%%%%%%%%%%%%%%%%%%%%%%%%%%%%%%%%

\rightline{LAPTH-Conf-054/12}

\bigskip

\textbf{\Large{Coordinate Bethe Ans\"atze for  non-diagonal boundaries}}\\[1.2ex]

{\large Eric Ragoucy\footnote{email: ragoucy@lapth.cnrs.fr}}

{\it Laboratoire de Physique Th\'eorique LAPTH,

 CNRS and Universit\'e de Savoie, BP 110, 74941 Annecy-le-Vieux Cedex, France}

\bigskip 

Talk given at \textit{Quantum Integrable Systems and Geometry} (3-7 September 2012, Olhao, Portugal) and based on joint works with S. Belliard, N. Cramp\'e and D. Simon, see 
\texttt{arXiv:1009.4119}, \texttt{arXiv:1105.4119}, \texttt{arXiv:1106.3264} and \texttt{arXiv:1209.4269}.

This short note presents only the basic ideas of the technique, and does not attend to give a general overview of the subject. Interested readers should refer to the original papers and references therein.

\bigskip

Bethe ansatz goes back to 1931, when H. Bethe invented it to solve some one-dimensional models, such as XXX spin chain, proposed by W. Heisenberg in 1928. Although it is a very powerful method to compute eigenvalues and eigenvectors of the corresponding Hamiltonian, it can be applied only for very specific boundary conditions: periodic boundary ones, and so-called open-diagonal boundary ones. After reviewing this method, I will present a generalisation of it that applies also to open-triangular boundary conditions.

%%%%%%%%%%%%%%%%%%%%%%%%%%%%%%%%%%%%%%%

\section{XXX model with periodic boundary conditions}
This model \cite{Heisen} describes the  interaction of spin $\half$ on a 1d lattice (of $L$ sites). The interaction is between nearest neighbours, and we assume the periodicity condition $L+1\equiv 1$ (periodic boundary conditions, also called closed chain).
\paragraph{Hamiltonian.}
Since we have $L$ spin $\half$, the total space of states
(Hilbert space) is $\cH=\big(\CC^2\big)^{\otimes L}$.

An arbitrary state in $\cH$ as the form
$$
\underbrace{|\up\rangle\otimes|\up\rangle\otimes|\down\rangle\otimes\cdots
 |\up\rangle\otimes|\down\rangle}_L	=|\up\up\down\ldots\up\down\rangle\,.
$$
 
The Hamiltonian is an operator acting on $\cH$:
\beano
H=\sum_{\ell=1}^{L} h_{\ell,\ell+1}=\sum_{\ell=1}^{L-1} h_{\ell,\ell+1}\ 
{+h_{L,1}}\in\mbox{End}(\cH)\,.
\eeano

The precise form of the "local" Hamiltonians $h_{\ell,\ell+1}$ is given below, but before coming to that point we need to specify some notations.

\paragraph{Notation.}

Indices indicate on 
which sites (of the chain) operators act non trivially. 
For instance 
$h_{34}=\II\otimes\II\otimes\,h\,\otimes(\II)^{\otimes 
(L-4)}\in\mbox{End}(\cH)$, where
$h$ is the "local" Hamiltonian acting on 2 sites, $h\in\mbox{End}(\CC^2\otimes \CC^2)$.

For the XXX model, the exact form of the local Hamiltonian is 
$
h_{\ell,\ell+1} = P_{\ell,\ell+1}-\II
$,
 where $P_{\ell,\ell+1}$ is the 
permutation operator acting on site $(\ell,\ell+1)$:
\beano
P&=&\left(\begin{array}{cccc} 1 & 0 & 0 & 0 \\ 0 & 0 & 1 & 0 \\
0 & 1 & 0 & 0 \\ 0 & 0 & 0 & 1 \end{array}\right)
\quad\Rightarrow\quad P\,( {u}\otimes {v}\,) = {v}\otimes {u}\,.
\eeano

Then, $h\, {u}\otimes {v}\, = {v}\otimes {u}-{u}\otimes {v}$, $\forall u,v= |\downarrow>,\,|\uparrow>$,
and in particular $h\,{u}\otimes {u} = 0$.

Note that $h$ does not change the total number of spin $\downarrow$ (or $\uparrow$). The problem we want to address is:
\textit{{What are spectrum and the eigenfunctions of $H$?}}
We aim at solving it through the coordinate Bethe ansatz.

%%%%%%%%%%%%%%%%%%%%%%%%%%%%%%%%%%%%%%%

\subsection{Coordinate Bethe ansatz}
Since we are looking for the Hamiltonian eigenfunctions $H\, \Phi = E\, \Phi$, $\Phi\in\cH$, we first remark that 
\beq
S^z=\sum_{j=1}^L \sigma^z_j
\mb{with} \sigma^z=\left(\begin{array}{cc} 1 & 0 \\ 0 & -1 \end{array}\right)
\eeq
commute with the Hamiltonian $[H\,,\, S^z]=0$, that is another way to say that the Hamiltonian does not change the total number of spin $\downarrow$ (or $\uparrow$). Then, we can decompose the Hilbert space into subspaces with fixed number of spin down: 
\beq\label{eq:Hm}
\cH=\oplus_{m=0}^L \cH_m\,,\quad \mbox{dim}\cH_m=\left(\begin{array}{c} L \\ m\end{array}\right)\,,\quad
S^z\,\vph_m=(L-2m)\vph_m\,,\quad \forall\vph_m\in\cH_m\,.
\eeq
It is enough to solve the eigenvalue problem in these subspaces:
\beq
\label{eq:sch1}
H\, \Phi_m = E\, \Phi_m\,,\quad \Phi_m\in\cH_m\,. 
\eeq
In $\cH_m$, a general state takes the form
\beano
&&|x_{1},x_2,\dots,x_m\rangle=
|\up\ldots\up\!\!\rouge{\raisebox{-1.ex}{$\atopn{\displaystyle\down}{\ x_{1}}$}}\!\!
\up\ldots\up\!\!\rouge{\raisebox{-1.ex}{$\atopn{\displaystyle\down}{\ x_{2}}$}}\!\!
\up\ldots\ldots\up\!\!
\rouge{\raisebox{-1.ex}{$\atopn{\displaystyle\down}{\ x_{m}}$}}\!\!\up\ldots\up\rangle
\in\big(\CC^2\big)^{\otimes L}
\\
&& 1\leq x_1< x_2<...<x_m\leq L\,.
\eeano
In this parametrization, the integers $x_j$'s refer to the position of the down spins  in a "sea" of up spins, and one has 
chosen as reference state the zero energy eigenvector in $\cH_0$:
$H\, |\up\ldots\up\rangle = 0$. Obviously, the opposite point of view (with up and down spins exchanged) can be taken, starting with $|\down\ldots\down\rangle$ and describing the position of up spins.

\medskip

To have a physical interpretation of the Bethe ansatz, it is worth looking for the eigenstates in $\cH_1$. A simple calculation shows that they take the form
\beq
\Phi_1(k)=\sum_{x=1}^L e^{ikx}\,|x\rangle \mb{with} e^{ikL}=1\,,\qquad H\,\Phi_1(k) = (e^{ik}+e^{-ik}-2)\,\Phi_1(k)\,.
\eeq
One recognizes in $\Phi_1(k)$, a plane wave  describing a spin down "excitation" (above the spin up "sea") that moves around the circle, with  momentum $k$. This momentum is quantized since we are in a finite dimensional system, and the different values $k_n=n\frac{2\pi}{L}$, $n=0,1,2,...,L-1$ give rise to $L=\mbox{dim}\cH_1$ different eigenstates.

\medskip

The Bethe Ansatz \cite{Bethe} assumes that in $\cH_m$, eigenstates can be constructed as superposition of $m$ planar waves, describing  $m$ spins down moving around the circle, with some momenta $(k_1,k_2,...,k_m)$. Obviously, since we are in a quantum system, one cannot attribute to a single spin down  a given momentum, and one needs to consider all the possible "configurations" between the momenta $(k_1,k_2,...,k_m)$ and the positions $(x_1,x_2,...,x_m)$. Then, the Bethe ansatz takes the form
\beano
 \Phi_m(k_1,...,k_m)=\sum_{x_{1}<\dots<x_m}\ \sum_{g\in S_m}\
A_g^{(m)}\ e^{i\boldsymbol{k}_g.\boldsymbol{x}}\ 
|x_{1},\dots,x_m\rangle\,, \mb{with} \boldsymbol{k}_{g}=(k_{g(1)},\dots,k_{g(m)}).
\eeano
 $S_m$ is the symmetric group ($A_{m-1}$ Weyl group), generated by  
transpositions $\ft_{j}$, $j=1,\ldots,m-1$ that exchange $k_j$ 
and $k_{j+1}$.
The coefficients $A_{g}^{(m)}$ are complex numbers to be determined 
such that (\ref{eq:sch1}) is satisfied.

\bigskip

We project equation (\ref{eq:sch1}) on the different independent 
vectors ${|x_{1},\dots,x_m\rangle}$ to get constraints on the coefficients 
$A^{(m)}_{g}$:

\begin{itemize}
\item  all the $x_{j}$'s are far away one from each other 
($1+x_j<x_{j+1}$, $\forall\,j$) and are not on the 
boundary sites 1 and $L$. This case will be called 
{generic}.
\item $x_{j}+1=x_{j+1}$ for one given $j$  (the remaining $x_\ell$ being generic),
\item $x_{m}=L$  (the remaining $x_\ell$ being generic). One could choose equivalently $x_{1}=1$ because of the
periodicity condition $L+1\equiv 1$.
\end{itemize}
{By linearity, more complicated cases just appear as superposition of  
`simple' ones.}

%%%%%%%%%%%%%%%%%%%%%%%%%%%%%%%%%%%%%%%
\paragraph{Calculation of the energy: projection on 
$\boldsymbol{|x_{1},\dots,x_m\rangle}$ generic.}
It gives the spectrum as a function of momenta $k_j$
\beano
E_m=\sum_{j=1}^m(e^{ik_j}+e^{-ik_j}-2)\,.
\end{eqnarray*}

\paragraph{Scattering matrix: projection on 
$\boldsymbol{|x_{1},\dots,x_{j},x_{j+1}=1+x_{j},\dots,x_m\rangle}$.}

It provides the scattering matrix between pseudo-excitations. 
\begin{eqnarray*}
A^{(m)}_{g\ft_j}
&=&S\!\left(e^{ik_{gj}},e^{ik_{g(j+1)}}\right)\,A^{(m)}_{g}\,,
\mb{with}
{S(z_1,z_2)}=-\frac{2z_2-z_1z_2-1}{2z_1-z_1z_2-1}\,.
\end{eqnarray*}
Since the symmetric group is generated by the transpositions $\ft_j$, the above equation allows to reconstruct $A^{(m)}_{g}$, $g\in S_m$, in term of a single one, say $A^{(m)}_{id}$. Then, one knows
 the wave function $\Phi_m$, as a function of momenta $k_j$.

\paragraph{Bethe equations: projection on $\boldsymbol{|x_{1}\dots,x_{m-1},L\rangle}$.} 
This last case leads to a series of equations for the momenta $k_j$:
\begin{eqnarray*}
&& \prod_{\substack{\ell=1 \\ \ell\neq j}}^m 
S(e^{ik_\ell},e^{ik_j})=e^{iLk_j}
 \mb{for} 1\leq j \leq m\,.
\end{eqnarray*}
In this way, we get a quantization of the pseudo-excitations momentas $k_{j}$, which is not surprizing 
since the system is in a finite volume. These equations are called the Bethe equations.

\medskip

Hence, we have obtained the spectrum and the eigenstates as functions of momenta that have been quantized: one could say that the problem is solved. Of course, one should also solve 
 the Bethe equations, but the full analytical resolution is not known yet. However, one can already perform a lot  (as the calculation of some correlation functions) just using the form of the Bethe equations. Moreover, the problem is numerically simpler to solve, than the diagonalisation of the full original Hamiltonian.

%%%%%%%%%%%%%%%%%%%%%%%%%%%%%%%%%%%%%%%

\section{XXX model with boundaries\label{sec:modXXX}}
We wish now to apply the same technique to the case of an "open XXX model", i.e. a XXX model defined not a circle anymore, but on a segment. In other words, the periodicity condition has been dropped out and replaced by some boundary conditions at site 1 and site $L$ in the following way. The
 Hamiltonian reads
\beano
H={B_{1}^+}\, +H_{bulk}+{B_{L}^-} \mb{where} 
H_{bulk}=\sum_{\ell=1}^{{L-1}} h_{\ell,\ell+1}
=\sum_{\ell=1}^{{L-1}} \Big(P_{\ell,\ell+1}-\II\Big)
\eeano
with boundary matrices
\beano
B^+=M^{-1}\,\left(\begin{array}{cc} \alpha & {\mu} \\ 0 & \beta 
\end{array}\right)\,M
\mb{and} B^-=M^{-1}\,\left(\begin{array}{cc} \gamma & 0 \\ 0 & \delta 
\end{array}\right)\,M\,,
\end{eqnarray*}
where $M$ can be any invertible 2 x 2 matrix. From the $SU(2)$-invariance of the local Hamiltonian 
$$M_1 M_2 h_{12} M^{-1}_1 M^{-1}_2 = h_{12}\,, $$
one can consider $M_1M_2...M_L\, H\, (M_1M_2...M_L)^{-1}$ instead of $H$ itself, so that it is enough to work out the case $M=\II$. We focus on this case in the following.

The Hamiltonian describes the interaction of spins 
(up or down) among themselves, and with two boundaries described by 
the matrices $B^\pm$. 

In general, one can perform the coordinate Bethe ansatz when the two boundary matrice $B^\pm$ are diagonal (i.e. when $\mu=0$). We will present a modification of the ansatz that can deal with triangular matrices. Before coming to this point, we present the usual ansatz, done for $\mu=0$.

\subsection{"Usual" Coordinate Bethe Ansatz}

When $\mu=0$ (diagonal boundaries), the boundaries do not 
modify the spin (no flip $\up$ to $\down$ or vice-versa). 
Then, the Hamiltonian, still preserve the number of spin down and we can still look for eigenstates in the 
spaces $\cH_m$. 
One can use the "usual" coordinate Bethe ansatz:
\beano
 \Phi_m=\sum_{x_{1}<\dots<x_m}\ \sum_{{g\in W\!B_m}}\
A_g^{(m)}\ e^{i\boldsymbol{k}_g.\boldsymbol{x}}\ 
|x_{1},\dots,x_m\rangle\,,
\end{eqnarray*}
However, since their are boundaries, the plane wave bounce back on them, so that when considering a momentum $k$, one needs to consider also the momenta $-k$. This is why the summation is not done on the symmetric group anymore, but on $W\!B_m$, the $B_{m}$ Weyl group. $W\!B_m$ is  generated by  
the symmetric group $S_m$ and the reflexion $R_{1}$ exchanging $k_{1}$ and 
$-k_{1}$.

Apart from this change, the technique stays the same, and the coefficients $A_{g}^{(m)}$ are all determined (but one) by  $ H \,\Phi_m = E_m\, \Phi_m$.
We project this equation on states $|x_{1},x_2,...,x_m\,\rangle$ with:
\begin{itemize}
\item  $(x_{1},x_2,...,x_m)$  generic ($1+x_j<x_{j+1}$, $\forall\,j$) 
\item $x_{j}+1=x_{j+1}$ for some $j$ (the remaining $x_\ell$ being generic)
\item $x_{m}=L$  (the remaining $x_\ell$ being generic)
\item $x_{1}=1$  (the remaining $x_\ell$ being generic)
\end{itemize}
Remark that now the cases $x_1=1$ and $x_m=L$ have to be considered separately, since there is no periodicity.

\paragraph{Calculation of the energy: projection on 
$\boldsymbol{|x_{1},\dots,x_m\rangle}$ generic.}
The calculation is similar to the periodic case, and we get
\beano
E_m=\alpha+\gamma+\sum_{j=1}^m(e^{ik_j}+e^{-ik_j}-2)\,.
\end{eqnarray*}
One can see the same "bulk" part as in the periodic case, plus a contribution from the two boundaries.

\paragraph{Scattering matrix: projection on 
$\boldsymbol{|x_{1},\dots,x_{j},x_{j+1}=1+x_{j},\dots,x_m\rangle}$.}
Again, the calculation is similar to the periodic case (boundaries are not involved in this process), and we get the same scattering matrix:
\begin{eqnarray} 
A^{(m)}_{g\ft_j} &=&
S\!\left(e^{ik_{gj}},e^{ik_{g(j+1)}}\right)\,A^{(m)}_{g}
\mb{with}
 {S(z_1,z_2)}=-\frac{2z_2-z_1z_2-1}{2z_1-z_1z_2-1}\,.
 \label{eq:scatt}
\end{eqnarray}
Note however that the above relation is not enough to reconstruct the whole eigenstates, because the transpositions $\ft_j$ are not enough to generated the $W\!B_m$ Weyl group. One needs another projection to be able to reconstruct the whole group.

\paragraph{Reflection coefficient for the left boundary: projection on 
$\boldsymbol{|1,x_{2}\dots,x_m\rangle}$}
This case is new for open case and characterizes the presence of a boundary. 
\bea
&&A^{(m)}_{gR_{1}} = R_+(e^{ik_{g1}})\ A^{(m)}_{g}
\mb{with} R_+(z)=-z^2\,\frac{1-\frac1z+\beta-\alpha}{1-z+\beta-\alpha}\equiv \frac{r_+(\frac1z)}{r_+(z)}\,.
 \label{eq:refl}
\end{eqnarray}
Equations (\ref{eq:scatt}) and (\ref{eq:refl}) are now sufficient to reconstruct all the coefficients $A^{(m)}_g$, $g\in W\!B_m$. We choose the site 1 to compute the reflection coefficient on this site, but obviously one can do the same calculation starting with site $L$. One needs only one reflection coefficient to recontruct the eigenstates, but both are needed to get the Bethe equations.

\paragraph{Bethe equations: projection on 
$\boldsymbol{|x_{1}\dots,x_{m-1},L\rangle}$} 
\begin{eqnarray*}
&& \prod_{\substack{\ell=1 \\ \ell\neq j}}^m
S(e^{ik_\ell},e^{ik_j})\,S(e^{-ik_j},e^{ik_\ell})
=e^{2iLk_j}\,
\frac{r_+(e^{ik_{j}})\,r_-(e^{ik_{j}})}
{r_+(e^{-ik_{j}})\,r_-(e^{-ik_{j}})} \,,
\qquad 1\leq j \leq m
\qquad\\
&&r_+(z) = \frac{(z-1)(1-z+\beta-\alpha)}{z(1+z)}
\mb{and} r_{-}(z) = \frac{z-1}{z+1}\,(1-z+\delta-\gamma)\,.
\end{eqnarray*}
Note that the two reflection coefficients (corresponding to boundaries 1 and $L$) are involved in the equation, restoring the apparent dissymmetry in choosing site 1 for the reflection coefficient.

\subsection{New case: $B^+$ is triangular}

When $\mu\neq0$, the left boundary can now flip the spin $\down$ to a spin $\up$. This means that the number of spins down is not conserved anymore, and we cannot use the decomposition (\ref{eq:Hm}) to find the Hamiltonian eigenstates.
However, since a spin up never can turn down, the \textit{maximal} number of spin down in a state is a well-defined quantity. This leads to the following modification of the ansatz:
\beano
\Psi_n(k_1,k_2,...,k_n) &=& {\sum_{m=0}^n}\ {\sum_{x_{m+1}<\dots<x_n}\ \sum_{{g\in G_m}}\
A_g^{(n,m)}\ e^{i\boldsymbol{k}^{(m)}_g.\boldsymbol{x}^{(m)}}\ |x_{m+1},\dots,x_n\rangle}\,,
\\
\boldsymbol{k}^{(m)}_g.\boldsymbol{x}^{(m)} &=&
k_{g(m+1)} x_{m+1}+k_{g(m+2)} x_{m+2}+...+k_{g(n)} x_{n}
\end{eqnarray*}
One can see that roughly speaking, the new ansatz has the form $ \sum_{m=0}^{n} \Phi_m$, where $\Phi_m$ denote the "original" anstaz: apart from the top component (corresponding to $m=0$), we added a "tail" of lower component corresponding to $m=1,...,n$ spins down that have "dropped" into the left boundary. 
Note also that the Weyl group $W\!B_n$ has been replaced by the coset
$G_m=W\!B_n/W\!B_m$. Indeed when $m$ spins down have dropped into the boundary, one has to distribute the $n$ momenta $k_j$ to only $(n-m)$ excitations, hence the use of the coset $G_m$. 

The coefficients $A_{g}^{(n,m)}$, $0\leq m\leq n$, are determined by  the equation
$ H \,\Psi_n = E_n\, \Psi_n$.
We project this equation on states $|\vec{x}\,\rangle$ with:
\begin{itemize}
\item  $(x_{1},x_2,...,x_n)$  generic ($1+x_j<x_{j+1}$, $\forall\,j$) 
\item $x_{j}+1=x_{j+1}$ for some $j$
\item $x_{n}=L$
\item $x_{1}=1$
\item $(x_{m+1},...,x_n)$ generic ($m>0$)
\end{itemize}

\paragraph{Calculation of the energy: projection on $\boldsymbol{|x_{1},\dots,x_n\rangle}$ generic.}
Here we are projecting on a state that corresponds only to the top component: the result is the same as for the usual ansatz.
\beano
E_n=\alpha+\gamma+\sum_{j=1}^n\lambda(e^{ik_j})
\mb{where}
\lambda(z)=z+\frac{1}{z}-2=\frac{(z-1)^2}{z}\,.
\end{eqnarray*}

\paragraph{Scattering matrix: projection on 
$\boldsymbol{|x_{1},\dots,x_{j},x_{j+1}=1+x_{j},\dots,x_n\rangle}$.}
Again, only the top component is involved, and we get the same result as for the periodic case:
\begin{eqnarray*} 
A^{(n,0)}_{g\ft_j} &=&
S\!\left(e^{ik_{gj}},e^{ik_{g(j+1)}}\right)\,A^{(n,0)}_{g}\,,
\mb{with}
{S(z_1,z_2)} = -\frac{2z_2-z_1z_2-1}{2z_1-z_1z_2-1}\,.
\end{eqnarray*}
This relation allows to reconstruct \textit{some} of the coefficients $A^{(n,0)}_g$.

\paragraph{Reflection coefficient for the left boundary: projection on 
$\boldsymbol{|1,x_{m+1}\dots,x_n\rangle}$.}
As for the diagonal boundaries case, this projection (together with the previous one) 
allows to reconstruct  the coefficients $A^{(n,0)}_g$ in term of say $A^{(n,0)}_{id}$:
$$
A^{(n,0)}_{gR_{1}} = R_+(e^{ik_{g1}})\ A^{(n,0)}_{g}
\mb{with}R_+(z)=-z^2\,\frac{1-\frac1z+\beta-\alpha}{1-z+\beta-\alpha}\,.
$$
However, we need more relations to obtain the coefficients $A^{(n,m)}_g$, $m>0$, that are in the "tail" of $\Psi_n$.

\paragraph{Transmission coefficient: projection on 
$\boldsymbol{|x_{m+1}\dots,x_n\rangle}$.}
This type of projection is new for triangular boundary matrices:
\beano
A^{(n,m)}_{g} &=& T^{(m)}(e^{ik_{g1}},...,e^{ik_{gm}})\ A^{(n,m-1)}_{g}\,,
\\
T^{(m)}(z_1,...,z_m)
&=&\frac{{\mu}}{r_+(z_m)\,\prod_{j=1}^{m-1}a(z_{m},z_j)\,a(z_{j},1/z_m)}\,,
\\
a(z_1,z_2) &=& i\,\frac{2z_2-z_1 z_2-1}{z_1 z_2-1}\mb{;}
r_+(z) =-\frac{(z-1)(1-z+\beta-\alpha)}{z(1+z)}\,.
\end{eqnarray*}
The relation allows to reconstruct all the coefficients $A^{(n,m)}_g$, $m>0$, in term of $A^{(n,0)}_{id}$. Thus, we get the complete form of $\Psi_n$. Remark that the transmission coefficient $T^{(m)}(z_1,...,z_m)$ is proportional to $\mu$: for diagonal boundary matrix ($\mu=0$), one recovers that the "tail" of $\Phi_n$ is not needed. The top component is already an eigenstate.

\paragraph{Bethe equations: projection on 
$\boldsymbol{|x_{1},\dots,x_{n-1},L\rangle}$.} 

\begin{eqnarray*}
&& \prod_{\substack{\ell=1 \\ \ell\neq j}}^n 
S(e^{ik_\ell},e^{ik_j})\,S(e^{-ik_j},e^{ik_\ell})
=e^{2iLk_j}\,
\frac{r_+(e^{ik_{j}})\,r_-(e^{ik_{j}})}
{r_+(e^{-ik_{j}})\,r_-(e^{-ik_{j}})} \,,
\qquad\ 1\leq j \leq n
\quad\\
&& r_{-}(z) = \frac{z-1}{z+1}\,(1-z+\delta-\gamma)\,.
\end{eqnarray*}
Again, the symmetry between left and right boundaries is restored in the Bethe equtions.

\section{Generalization to XXZ model with boundaries}
The same type of generalized ansatz can be done for a more complicated system, the XXZ model.
\subsection{Hamiltonian of the model}
The XXZ Hamiltonian takes the form
\beano
H_{XXZ} &=&  -\half\sum_{j=1}^{L-1}\Big\{\sigma^x_j\sigma^x_{j+1}
+\sigma^y_j\sigma^y_{j+1}+\Delta(\sigma^z_j\sigma^z_{j+1}-\II)-\fh\,(\sigma^z_j-\sigma^z_{j+1})\Big\}
+\wh B_1 + B_L\,,
\label{eq:hamxxz}\\
\Delta &=& \frac12(Q+Q^{-1}) \mb{and} \fh\,=\, \frac12(Q-Q^{-1})\,, 
%\mb{and} Q=\sqrt{\frac qp}
\\
\widehat B &=&
\left(\begin{array}{c c}
\alpha  & -\gamma e^{-s}  
\\
 -\alpha e^{s}  & \gamma
\end{array}\right)
\mb{and}
B =
\left(\begin{array}{c c}
\delta  & -\beta Q^{L-1}  
\\
-\frac{\delta}{Q^{L-1}} & \beta
\end{array}\right)\,.
\label{eq:Bxxz}
\eeano
Apart from $Q$, the bulk parameter corresponding to the deformation parameter of the underlying quantum group structure, there are five 
boundary parameters: $\alpha, \beta, \gamma, \delta$ and $s$ that enter into the two boundary matrices of the model.

\subsection{Generalized Coordinate Bethe Ansatz\label{sec:gCBA-XXZ}}
Since the technique is roughly the same, we just  emphasize the 
 extra new features that occur.

\paragraph{Constraints are imposed on the boundary parameters.} 

They take the form introduced in the original approach on non-diagonal boundary matrices \cite{Nepo}
\bea
&&\prod_{\eps,\eps'=\pm}\Big(c_\eps(\alpha,\gamma)\,c_{\eps'}(\beta,\delta)-Q^{L-1-n}\,e^{-s}\Big)=0
\mb{with} 0\leq n\leq L-1\label{eq:c1-1}\\
&&\mb{where} c_+(z_1,z_2)=\frac {z_1}{z_2} \  \mbox{and}\  c_-(z_1,z_2)=1\,.
\eea
One term of the above product must be zero, so that a given choice of constrains correspond to a triplet $(n,\eps,\eps')$. Note that another set of constraints is also available, see \cite{XXZ}.

\paragraph{Basis vectors depend on which site they are.} 

This change of basis corresponds to the local gauge transformations \cite{gauge}
\begin{eqnarray}
|\up\rangle_i \ \to\ |u\rangle_i= \begin{pmatrix} 1 \\  {Q^{1-i}}\,u \end{pmatrix}_i
\quad;\quad
|\down\rangle_i \ \to\ |d\rangle_i= \begin{pmatrix} 1 \\  {Q^{1-i}}\,d\end{pmatrix}_i\,,\qquad
i=1, 2,...,L\,.\label{base}
\end{eqnarray}
The vectors depend also on a free parameter ($u$ or $d$) that are used in the ansatz. Once this change of basis is done, one constructs generic states of $\cH$ as:
\begin{eqnarray}\label{eq:vecx}
|x_{m+1},\ldots,x_n\rangle=
|u_{m+1}\ldots u_{m+1} {\raisebox{-0.81ex}{$\atopn{\rouge{\displaystyle d_{m+1}}}{\rouge{x_{m+1}}}$}} 
u_{m+2}\ldots u_{n}
{\raisebox{-.81ex}{$\atopn{\rouge{ \displaystyle d_{n}}}{\rouge{x_{n}}}$}}u_{n+1}\ldots u_{n+1}\rangle\,,
\end{eqnarray}
where the parameters $u_\ell$ and $d_\ell$, $\ell=m+1,...,n$, are related by the relations $u_\ell=Qu_{\ell-1}$ and $d_\ell=Qd_{\ell-1}$, in addition to the rule given in (\ref{base}).

\paragraph{Telescoping terms appear locally.}
One introduces an auxiliary vector $|t\rangle =\left(\begin{array}{c} Q^{-1}-Q \\ 0 \end{array}\right)$ that allows a simple expression for the action of local Hamiltonians:
\begin{eqnarray}
h_{12} |u\rangle\otimes |u\rangle &=&  0 \mb{ ; }
h_{12} |d\rangle\otimes |d\rangle =    |t\rangle\otimes |d\rangle - |d\rangle\otimes |t\rangle \,,
\\
h_{12} |d\rangle\otimes |u\rangle &=&  Q^{-1} \, |u\rangle\otimes |d\rangle - |d\rangle\otimes |u\rangle -
  |d\rangle\otimes |t\rangle \,,\label{eq:hdu}
\\
h_{12} |u\rangle\otimes |d\rangle &=&  Q\,  |d\rangle\otimes |u\rangle - |u\rangle\otimes |d\rangle 
 +|u\rangle\otimes |t\rangle\,. \label{eq:hud}
\end{eqnarray}
Remark the opposite signs in (\ref{eq:hdu}) and (\ref{eq:hud}), so that the telescoping terms
 cancel in the bulk $\sum_j h_{j,j+1}$ and are used on the boundaries to diagonalize them.

\medskip

Apart from these "technicalities" the ansatz is the same:
\beano
\Psi_n= {\sum_{m=0}^n}\ {\sum_{x_{m+1}<\dots<x_n}\ \sum_{{g\in G_m}}\
A_g^{(n,m)}\ e^{i\boldsymbol{k}^{(m)}_g.\boldsymbol{x}^{(m)}}\ 
|x_{m+1},\dots,x_n\rangle}\,,
\end{eqnarray*}
with as for the XXX model, $G_m=W\!B_n/W\!B_m$ and $\boldsymbol{k}^{(m)}_g.\boldsymbol{x}^{(m)}=
\sum_{j=m+1}^n k_{gj} x_j$, but remember that the vector $|x_{m+1},\dots,x_n\rangle$ has now the form 
(\ref{eq:vecx}).

Then, the resolution follows the same lines and provide spectrum, eigenfunctions and quantization of rapidities $k_j$.
However note that $n$ in $\Psi_n$ is fixed and 
 given by the choice of constraint $(n,\eps,\eps')$. It appears to be the maximum number of "excitations" in $\Psi_n$. 
Then, it is clear that 
the present ansatz do \underline{not} provide all the spectrum. 

\paragraph{Matrix Coordinate Bethe ansatz.}
One needs to use another ansatz to get a complete spectrum.  The Matrix ansatz \cite{MA} (used in non-equilibrium statistical physics) gives another eigenstate (with zero eigenvalue), but in general it is not sufficient. To get the full spectrum, we 
 developed Matrix Coordinate Bethe ansatz, that can be viewed as 
a mixing of generalized coordinate Bethe ansatz and of Matrix ansatz. 
It can be also viewed as a non-commutative generalized coordinate Bethe ansatz. 
It uses  an extra (auxiliary) algebra with two generators $E$ and $D$ submitted to $qED-pDE = D+E $ (with $Q=\sqrt{q/p}$) and its representations. Essentially, the algebra is "hidden" in the definition of the basis vectors (\ref{eq:vecx}), as for the local gauge transformations used in section \ref{sec:gCBA-XXZ}, and the constraints (\ref{eq:c1-1}) select which type of representation one needs to use for the ansatz. 

Let us stress that this new ansatz is not an alternative way of solving the model but is rather \textit{complementary} to the generalized coordinate Bethe ansatz. Indeed, it provides a \textit{different} set of eigenvalues of the Hamiltonian, so that to get the full spectrum, one needs \textit{both} ans\"atze (generalized \textit{and} Matrix coordinate Bethe ansatz). One can say that they are based on two different reference states (the so-called pseudo-vaccua) that cannot be related through the ansatz.
For more details see \cite{matrix}.

\section{Conclusion: Open questions}
We have performed the coordinate Bethe ansatz in the case of XXX and XXZ models with non-diagonal boundary matrices (submitted to some constraints). We believe that this result can give new insight to the resolution of models with non-diagonal boundary matrices. Of course, to work out correlation functions for these models, an algebraic version of the Bethe ansatz for such models should be produced. Indeed, using the technique developed for the coordinate Bethe ansatz, we performed the algebraic Bethe ansatz in the case of XXX model with upper triangular matrices \cite{algebraic}. The case of XXZ model remains to be done. The case of higher rank algebras should also be considered.

Concerning the XXZ models, two types of ans\"atze were needed to get the full spectrum: a unified version of  these two ans\"atze has to be done. We believe it could give some hint for the case of fully non-diagonal boundary matrices (with no constraint).

\end{document}